\documentclass[%
 aip,
 amsmath,amssymb,
 reprint,
]{revtex4-2}

\usepackage{graphicx}
\usepackage{dcolumn}
\usepackage{bm}

\usepackage[utf8]{inputenc}
\usepackage{mathptmx}
\usepackage{etoolbox}

\makeatletter
\def\@email#1#2{%
 \endgroup
 \patchcmd{\titleblock@produce}
  {\frontmatter@RRAPformat}
  {\frontmatter@RRAPformat{\produce@RRAP{*#1\href{mailto:#2}{#2}}}\frontmatter@RRAPformat}
  {}{}
}%
\makeatother
\begin{document}


\title{Thermal hysteresis and the heat shuttling effect}

\author{Jean-Claude Krapez}
 \email{jean-claude.krapez@onera.fr}
\affiliation{%
 ONERA, The French Aerospace Lab, DOTA, F-13661 Salon cedex Air, France
}%






\begin{abstract}
Phononics has attracted much attention driven by the promising potentials offered by devices such as thermal diodes, thermal transistors, and thermal memristors. Heat shuttling (or heat ratcheting, or heat pumping) is a phenomenon exhibited by nonlinear materials presenting temperature-dependent thermal conductivity which, when sandwiched between two thermal baths 
with one bath subjected to a time-varying temperature, show non vanishing net heat flow, although the baths share the same average temperature. Phase-change materials (PCMs) like VO$_{2}$ were recently taken for illustration due to a strong change in conductivity over a small temperature range; energy extraction from the thermal variations of the environment was envisioned thereupon. However, up to now, the impact of PCM hysteresis has been overlooked or underestimated. On the basis of a thermal model simulating partial hysteresis loops and non-hysteretic branches, we demonstrate that the presence of hysteresis profoundly modifies the appearance of the heat shuttling effect and can constitute a hindrance to its manifestation. Operating configurations to improve its observation have been proposed.

\end{abstract}

\maketitle


\section{\label{sec:Introduction}Introduction}
Controlled manipulation of heat flow is of high importance for the development of a wide variety of technologies. Nonlinear heat transfer is essential to tailor heat currents; for this purpose, one can leverage the variations of thermal conductivity or emissivity with temperature. This thermal dependence generates an asymmetric system response that has been used to manipulate heat flows carried by phonons (phononics) and photons (photonics) in the same manner as the flow of electrons is controlled in electronic circuits. This gave rise to applications both at nanoscale \cite{li2012colloquium} and at macroscopic scale \cite{dai2021designing}, among them thermal diodes \cite{li2004thermal,terraneo2002controlling,ordonez2018conductive,schmotz2011thermal}, thermal transistors \cite{li2006negative,li2012colloquium, ordonez2016transistorlike, latella2019dynamical}, thermal logic gates \cite{wang2007thermal}, thermal memories \cite{wang2008thermal}, and thermal memristors  \cite{ben2017thermal, yang2019theoretical}.
When submitted to temporal modulation, devices supporting nonlinear heat flow may give rise to novel and intriguing phenomena, among them heat shuttling, a net heat current that shows up even in the absence of a mean temperature gradient. The shuttling of heat (or heat ratcheting, or heat pumping) has been discussed a few years ago in the context of heat conduction in nonlinear lattices \cite{li2008ratcheting, li2009shuttling, ren2010emergence}. A salient requirement is the presence of an induced dynamical symmetry-breaking mechanism in conjunction with nonlinearity \cite{li2012colloquium}. The necessary symmetry-breaking can be obtained by using two coupled lattices with different thermal properties \cite{li2008ratcheting, ren2010emergence} or by exploiting the nonlinear response induced by the harmonic mixing mechanism stemming from a time varying two-mode modulation of the bath temperature \cite{li2009shuttling}. Later on, the shuttling effect has been demonstrated for the radiative heat flux exchanged between two bodies separated by a thin vacuum gap \cite{latella2018radiative}. Recently, heat shuttling obtained by heat conduction at the macroscopic scale in phase change materials (PCMs) has been discussed \cite{liu2022energy,ordonez2022net}. The heat shuttling phenomenon was explained by the existence of a temperature-dependent thermal conductivity: when conductivity increases with temperature, a net heat conduction flow is observed from the bath whose temperature is symmetrically modulated to the static bath, whereas the opposite is observed when conductivity decreases with temperature \cite{ordonez2022net}. The theoretical analysis performed by Ordonez et al. \cite{ordonez2022net} was illustrated by taking as an example the properties of VO$_{2}$, a PCM showing a strong increase of conductivity with temperature. The quantification of the heat shuttling effect was nevertheless made in the static (pseudo-modulated) regime while ignoring the hysteretic behavior of the material. The numerical modeling described by Liu et al. \cite{liu2022energy} focused on nitinol and VO$_{2}$ while restricting to the thermally-thick regime. In addition, the hysteresis regarding heat conductivity and specific heat was modeled by assuming an instantaneous switch from the heating curve to the cooling curve (and vice versa) after a change of the sign of the temperature evolution, which was recognized as a rough approximation \cite{liu2022energy}. 
In this paper we model the conductive heat shuttling effect in a PCM when considering all following phenomena : nonlinear conductivity, heat storage with a temperature-dependent heat capacity, or, equivalently, through nonlinear enthalpy, hysteresis regarding both conductivity and enthalpy. Incomplete phase change process will be modeled by applying to both conductivity and enthalpy the \emph{curve-switch} model first described by Bony and Citherlet \cite{bony2007numerical} for enthalpy alone. Illustrations for the case of VO$_{2}$ will be provided. We show that hysteresis has a strong impact on the net heat current: heat shuttling may totally disappear if the amplitude of the temperature modulation is lower than a threshold depending on the hysteresis width. We describe the optimal configuration for maximizing the heat shuttling effect taking as an example the thermal properties of VO$_{2}$. A parametric analysis regarding the modulation frequency will provide a description of the transitory behavior between the thermally thick and thermally thin regimes. The paper ends with an opening-up on a two-bath modulation offering a perspective for improving the heat shuttling effect.

\section{\label{sec:model}Thermal model}

\subsection{\label{sec:shuttling}Heat shuttling through temperature modulation of one thermal bath}
\begin{figure}[b]
\includegraphics[width=0.45\textwidth]{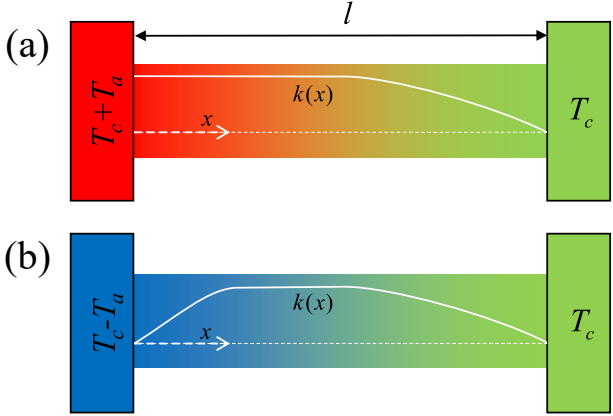}
\caption{\label{fig:fig_1} Scheme of a PCM, seat of heat shuttling by modulated heat conduction between two thermal baths. The right bath is at constant temperature $T_{c}$ whereas the left bath supports a periodic temperature variation of amplitude $T_{a}$ between $T_{c}+T_{a}$ (a) and $T_{c}-T_{a}$ (b). The color gradation reflects the temperature variations $T(x)$ while the white line describes the thermal conductivity profile $k(x)$ induced by the nonlinearity and the hysteresis (the absence of biunivocality between the profiles $T(x)$ and $k(x)$ is explained by the hysteresis).}
\end{figure}

Let us consider a system comprising a nonlinear material of thickness $l$ in thermal contact with two reservoirs (see Fig. \ref{fig:fig_1}). One of them (say the one at right) is at constant temperature $T_{c}$, the other one is at a (sinusoidally) modulated temperature with frequency $f$, amplitude $T_{a}$ and a mean equal to $T_{c}$:
 \begin{equation}\label{eq:BC_l}
T_{L}(t)=T_{c}+T_{a} \sin\left( 2\pi f t \right),
\end{equation}
 \begin{equation}\label{eq:BC_r}
T_{R}(t)=T_{c},
\end{equation}
for $t>0$. These conditions imply a zero mean gradient.

\subsection{\label{sec:properties}Thermal properties}
\begin{figure}[b]
\includegraphics[width=0.45\textwidth]{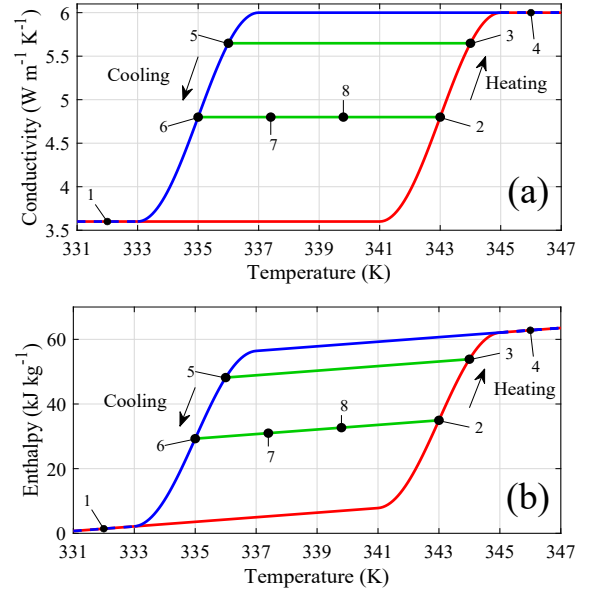}
\caption{\label{fig:fig_2} Evolution of the thermal conductivity (a)  and enthalpy (b) in relation to temperature during the heating (in red) and cooling (in blue) processes of VO$_2$. The green lines describe heating/cooling paths without phase change inside the major hysteresis loop. The numbers are used to describe full and incomplete hysteresis loops (see text).}
\end{figure}

As a nonlinear material we will consider VO$_{2}$, which, in addition to exhibiting a strong thermal conductivity change in a narrow range of temperature, presents a significant hysteresis. These phenomena are related to a phase change, more specifically a metal-insulator transition (MIT) in which the insulating and metallic phases coexist over a finite temperature range \cite{oh2010thermal}. At low temperature ($T<$ 330~K) VO$_{2}$ behaves like a dielectric with low thermal conductivity ($k_{i} = 3.6$~W~m$^{-1}$~K$^{-1}$), while at high temperature ($T>$ 345~K) it becomes a metal with a higher thermal conductivity ($k_{m} = 6$~W~m$^{-1}$~K$^{-1}$). Actually, the transition region is narrower, about 4~K. Because of hysteresis, a shift in the temperature dependence of the VO$_{2}$ properties is observed and has been reported to be about 8~K \cite{qazilbash2007mott}. According to this, we take for $T_{i}$ and $T_{m}$, the temperature-transition ends in the insulator state, resp. metallic state, as 341~K and 345~K during the heating process and as 333~K and 337~K during the cooling process.
The properties in the transition region can be described using an effective medium theory involving the volume fractions of the insulating and metallic regions and the respective properties \cite{qazilbash2007mott,ordonez2018modeling}. Here, to keep our theoretical description as simple as possible, we model the volume fraction $f$ of the metallic domains as a smooth step function of temperature given by:
\begin{equation}\label{eq:f}
f(T)=\frac{1}{2}\left(1-\cos(\pi \tilde{T})\right) \; ;\;
\tilde{T}=\frac{T-T_{i}}{T_{m}-T_{i}},
\end{equation}
and the conductivity as:
\begin{equation}\label{eq:k}
k(T)=k_{i}+(k_{m}-k_{i})f(T).
\end{equation}
Based on this model, the conductivity vs temperature curves are as reported in Fig. \ref{fig:fig_2}(a) for the heating (in red) and cooling (in blue) processes.

The values of the specific heat in the insulating and metallic phases are quite close and don't vary much in the short temperature range of $\sim$10~K resp. below and above the MIT that we intend to explore \cite{berglund1969electronic}. We thus considered a common and constant value of $C=0.71$~kJ~kg$^{-1}$~K$^{-1}$. The enthalpy per unit mass of VO$_{2}$ (actually sum of sensible enthalpy and latent heat) can thus be written as:
\begin{equation}\label{eq:H}
H(T)=C(T-T_{ref})+Lf(T),
\end{equation}
where $T_{ref}$ is an arbitrary reference temperature (here taken as 330~K) and $L=51.45$~kJ~kg$^{-1}$ is the latent heat of the transition \cite{berglund1969electronic}.
Thereby, the enthalpy values reached at the transition ends of the heating and cooling phases are, respectively, $\left\lbrace H_{i},H_{m} \right\rbrace =\left\lbrace 7.81,62.1 \right\rbrace $~kJ~kg$^{-1}$ and $\left\lbrace H_{i},H_{m} \right\rbrace=\left\lbrace 2.13,56.42 \right\rbrace $~kJ~kg$^{-1}$. The evolution of enthalpy with the temperature according to the present model is described in Fig. \ref{fig:fig_2}(b). During the numerical process, it will be necessary to extract the temperature from the current enthalpy value $H$, which amounts to apply an inversion to Eq.~(\ref{eq:H}). Three cases will be considered depending on how $H$ compares to the boundaries $H_{i}$ and $H_{m}$, whether in the heating or cooling phase:
\begin{equation}\label{eq:inv_H}
T(H)=\left\lbrace\begin{array}{lll}
T_{ref}+H/C & \;,\quad & H \leq H_{i}\\
T_{i}+\left(T_{m}-T_{i} \right)g(H) & \;,\quad & H_{i} \leq H \leq H_{i}\\
T_{ref}+\left(H-L\right)/C & \;,\quad & H \geq H_{m}
\end{array}\right.
\end{equation} 
where $g(H)$ is the inverse step function of $f(T)$. There is no closed-form analytical expression for $g(H)$, nevertheless, the following approximation gives satisfactory results:
\begin{equation}\label{eq:g}
g(H)=c_{1}\tilde{H}+\left(1- c_{1}\right) \left[1+\frac{h(1-\tilde{H})}{h(\tilde{H})}\right] ^{-1} ,
\end{equation}
with
\begin{equation}\label{eq:h}
h(x)=\left( 1+c_{2}x \right)^{c_{3}}-1,
\end{equation}
the normalized enthalpy
\begin{equation}\label{eq:H_tilde}
\tilde{H}=\left( H-H_{i}\right) / \left( H_{m}-H_{i}\right),
\end{equation}
and the empirical coefficients $c_1=0.211$, $c_{2}=5140$, $c_{3}=0.551$. The error on temperature is then less than $1.3\cdot 10^{-3}$K and the RMS error is $5\cdot 10^{-4}$K. The values of the three coefficients $c_1, c_{2}, c_{3}$ only depend on the value taken by the ratio of the sensible heat to latent heat $C(T_{m}-T_{i})/L$ (which is the same for both heating and cooling processes). 

\subsection{\label{sec:loops}Modeling the hysteresis loops}
Thermal cycling with complete phase changes from fully isolating phase to fully metallic phase and vice versa generates a full hysteresis loop which is described by the succession of points [1-2-3-4-5-6-1] in Fig. \ref{fig:fig_2}. The question is then: what happens if heating stops before the right end of the heating branch (i.e. 345~K) or, symmetrically, if cooling stops before the left end of the cooling branch (i.e. 333~K) ? This was described, for the electric-resistance versus temperature hysteresis, through the so-called first-order reversal curves \cite{ramirez2009first}. Incomplete hysteresis loops were also analysed for the electric-resistance and the optical reflectance by Gurvitch et al. \cite{gurvitch2009nonhysteretic}. In particular, temperature excursions taken from an attachment point on any side of the major hysteresis loop produce minor loops and, for sufficiently small excursions, these minor loops flatten out, degenerating into nonhysteretic branches (NHBs) that are linear in $log(\rho)$, or reflectance, versus $T$ \cite{gurvitch2009nonhysteretic}. The slope of these NHBs evolves from the resistance (or reflectance coefficient) observed in the semiconducting phase to the one in the metallic phase \cite{gurvitch2009nonhysteretic}. We will capitalize on these observations regarding NHBs and transpose them qualitatively to thermal conductivity. In addition, to keep our model sufficiently simple, we will consider that NHBs apply to minor loops of large excursions as well (i.e. up to the major loop width).
A few example will illustrate the possible trajectories of the representative point (RP) of VO$_2$ in the conductivity versus temperature map. We will admit that, if, after reaching a point along the heating branch as point 2 in Fig. \ref{fig:fig_2}(a) cooling starts, the conductivity stays constant while the RP moves along the horizontal (green) line until point 6 is reached. Then, if the cooling continues, the RP moves along the cooling branch towards point 1. Alternatively, cooling can stop along the green line at point 8 and heating resumes. Temperature and conductivity can then oscillate between any pair of points along the horizontal line 2-6.

The transition between points 2 and 6 is without phase change; the fraction of the metallic domains remains constant. Hence, enthalpy evolves according to the specific heat of the present semiconductor/metal mixture. Since we have assumed that the specific heat is the same for both phases in the narrow temperature range considered, the line joining points 2 and 6 (or 3 and 5) in Fig. \ref{fig:fig_2}(b) is a straight line with the same slope as that of the enthalpy curve to the left of point 1 and to the right of point 4. Advanced numerical models for heat transfer modeling in phase change materials used for energy storage in buildings apply this approach for evaluating the heat exchanges inside the enthalpy-hysteresis loop \cite{bony2007numerical,delcroix2017development}.

Examples of cyclic paths with incomplete phase changes are given in Fig.~\ref{fig:fig_2} by the cycles [1-2-6-1] (after initially heating to point 1), [4-5-3-4] (after initially cooling to point 4), and [2-3-5-6-2] (after initially heating from point 1 to point 2 or initially cooling from point 4 to point 5 and then joining the cycle). Examples of cyclic paths experiencing no phase change are given by the cycles [2-8-2] (after having initially heated from point 1 to point 2), [6-7-6] (after having initially cooled from point 4 to point 6), [2-6-2] (after having initially heated from point 1 to point 2 or cooled from point 4 to point 6), and [8-7-8] (by reaching the cycle through the paths 1-2-8 or 4-6-7).

\subsection{\label{sec:Numerical}Numerical model}
Assuming that the heat exchanges with the environment (radiation, convection) taking place all around the PCM, from one end to another, have a negligible impact, then the heat diffusion process is essentially one dimensional along the longitudinal direction $x$ and the heat equation to solve for the temperature $T(x,t)$ reduces to:
\begin{subequations}
\label{eq:Eq_heat}
\begin{equation}
\rho\frac{\partial H\left(T \left(x,t \right)\right)}{\partial t}=
-\frac{\partial }{\partial x}\left(\varphi \left(x,t \right) \right),\label{subeq:1}
\end{equation}
with the heat-conduction flux $\varphi(x,t)$ defined by:
\begin{equation}
\varphi(x,t)=
-k\left( T \left( x,t \right)\right) \frac{\partial T \left(x,t \right)}{\partial x},\label{subeq:2}
\end{equation}
\end{subequations}
where $\rho=4670$~kg~m$^{-3}$ is density of VO$_2$ \citep{leroux1998vo}, together with the Dirichlet boundary conditions in Eqs.~(\ref{eq:BC_l}) and (\ref{eq:BC_r}).
The resulting partial differential equation for $T(x,t)$ is both nonlinear and with memory since conductivity and enthalpy do not only depend on temperature but on whether it decreases or increases, see Fig.~\ref{fig:fig_2}. The problem was solved numerically by the finite volume method while implementing the enthalpy formalism \citep{swaminathan1993enthalpy} whose advantages make it a prefered tool for solving phase-change problems \citep{al2013modeling}. An explicit time-marching technique was adopted whose counterpart of simplicity is to constrain the time step to remain below an upper limit to grant numerical stability (Fourier criterion). For the treatment of combined hysteresis and partial phase change, the method of Bony and Citherlet \cite{bony2007numerical} for enthalpy has been applied and formally extented to conductivity, which formalizes the aspects discussed previously regarding the partial hysteresis loops (see also \cite{delcroix2017development,klimevs2020computer}). The details of the numerical model and its validation against analytical results in the case of linear diffusion can be found in the Supplementary Material. 

\section{\label{sec:Results}Results}

\subsection{\label{sec:modulations}Temperature and flux modulations}
To illustrate the heat shuttling effect in phase change materials with hysteresis, we will consider a 1~mm thick VO$_2$ layer. We first consider the particular case where the mean temperature of the modulated reservoir is equal to the half-rise temperature along the heating branch, namely $T_{c}=343$~K. Because of the memory effect which is a direct consequence of the hysteresis, it is important to clarify what the thermal history of the PCM has been \textit{before starting} the modulation scheme described by the boundary conditions in Eqs.~(\ref{eq:BC_l}) and (\ref{eq:BC_r}), i.e. for $t<0$. In this paper, we considered two different scenarios for the prior history. In the first one, the whole system is assumed initially at a temperature lower than the bottom temperature of the cooling branch ($T < 331$~K), after which the temperature of both reservoirs is increased to $T_{c}=343$~K. As a consequence, through thermal conduction, the PCM reaches (asymptotically) the same uniform equilibrium temperature $T_{c}$ \textit{from below}. Hence, at the starting point of the modulation, i.e. $t=0$, the coordinates $\left\lbrace T,\,k \right\rbrace$ of the RP in the hysteresis loop of any location inside the PCM are $\left\lbrace 343,\,4.8 \right\rbrace $, which corresponds to point n°2 in Fig.~\ref{fig:fig_2}(a). In the second scenario, the whole system is assumed initially at a temperature higher than the topmost temperature of the heating branch ($T > 345$~K), after which the temperature of both reservoirs is lowered to $T_{c}=343$~K. Similarly, the PCM reaches (asymptotically) the same uniform equilibrium temperature $T_{c}$, but \textit{from above}. Hence, the coordinates $\left\lbrace T,\,k \right\rbrace$ of the RP of any location inside the PCM are now $\left\lbrace 343,\,6 \right\rbrace $. In the second scenario, the temperature modulation of the left reservoir starts while the conductivity of the PCM is uniformly at 6~W$\,$m$^{-1}\,$K$^{-1}$, whereas it is 4.8~W$\,$m$^{-1}\,$K$^{-1}$ in the first scenario. 

\begin{figure}
\includegraphics[width=0.45\textwidth]{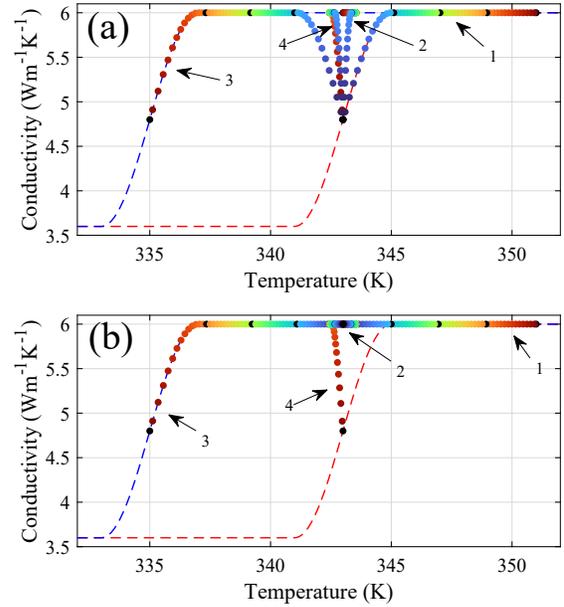}
\caption{\label{fig:fig_3}Distribution of the PCM discretization nodes in the temperature/conductivity space at four particular times during thermal cycling (i.e. after 1, 2, 3 and 4 quarters of a period). VO$_{2}$ sample thickness: 1 mm, frequency: 0.3 Hz, temperature amplitude: $T_{a}=8$~K, mean temperature $T_{c}=343$~K (i.e. the half-rise temperature along the heating branch). Before the onset of cycling, a thermal equilibrium at $T_{c}=343$~K is reached by increasing (a) or decreasing (b) the system temperature. The color dots, from red to blue, are representative of each numerical cell from left to right of the PCM (total of 53 cells). Black dots are used to designate points at $x=0,\,0.25l,\, 0.5l,\, 0.75l,\, l$. The heating branch of the VO$_{2}$ hysteresis loop is in red dashed line; the cooling branch is in blue dashed line (from Fig. \ref{fig:fig_2}(a)). See the videos of a whole cycle in the Supplementary Material in animation-fig-3a, resp. animation-fig-3b.}
\end{figure}

Transient effects are observed after the onset of the temperature modulation; they die out after a few periods. In this paper we will concentrate on the steady-state periodic response after the transient effects have vanished. The temperature distribution inside the PCM, at four particular times during a period (i.e. after 1, 2, 3 and 4 quarters of a period), is reported in Fig.~\ref{fig:fig_3}(a) for the first scenario, and in Fig.~\ref{fig:fig_3}(b) for the second scenario. In both cases, the modulation amplitude of the left reservoir is $T_{a}=8$~K and the frequency is $f=0.3$~Hz. At this frequency, the thermal diffusion length $(a/(\pi f))^{1/2}$, where $a$ is the thermal diffusivity, is about 1.4~mm in the metallic phase and 1.1~mm in the insulating phase, which are of the order of magnitude of the layer thickness. At this frequency the PCM layer is thus neither thermally thin nor thermally thick. Dots with graded colors were used to represent the position in the PCM of the cells of the numerical model (53 cells in the present case) for these four snapshots in the temperature/conductivity representation. Video animations of the evolution of the temperature/conductivity profiles during a whole period are provided in the Supplementary Material.

The representative point $ \left\lbrace T,\,k \right\rbrace$ of the right boundary remains stuck at $ \left\lbrace 343,\,4.8 \right\rbrace$ after the first prior scenario, resp. at $ \left\lbrace 343,\,6 \right\rbrace$ after the second prior scenario. Meanwhile, the representative point of the left boundary follows the same trajectory in the $ \left\lbrace T,\,k \right\rbrace$ plane for both prior scenarios (partial hysteresis loop). The values reached at the four particular times considered in Fig.~\ref{fig:fig_3} are $ \left\lbrace 351,\,6 \right\rbrace$, $ \left\lbrace 343,\,6 \right\rbrace$, $ \left\lbrace 335,\,4.8 \right\rbrace$ and $ \left\lbrace 343,\,4.8 \right\rbrace$ . This means that the left boundary has lost memory of the prior scenario. Moreover, only the left part of the PCM (somewhat less than one quarter of the thickness) sees its thermal conductivity evolve during the cycling. The RP follows a partial hysteresis loop that progressively flattens out as one moves away from the left reservoir. The remaining of the PCM layer shows temperature changes (of progressively lower amplitude while approaching the right boundary) \textit{but no conductivity changes} during the cycling (the representative point follows either a flat hysteresis loop or keeps in the full metallic phase). This means that less than 25\% of the PCM close to the left reservoir can manifest its nonlinearity and more than 75\% of the PCM close to the static reservoir, behaves either like a \textit{graded-conductivity linear} material after the 1\textsuperscript{st} prior scenario (see Fig.~\ref{fig:fig_1} which qualitatively illustrates the two extreme thermal distributions related to this case) or a \textit{homogeneous linear} material after the 2\textsuperscript{nd} prior scenario (the conductivity there remains uniformly equal to 6~W$\,$m$^{-1}\,$K$^{-1}$). The feared consequence is that the heat-shuttling effect, which has been previously described for a material exhibiting nonlinearity \textit{in the whole thickness} \citep{liu2022energy,ordonez2022net}, is attenuated in the curent situation. 

\begin{figure}
\includegraphics[width=0.45\textwidth]{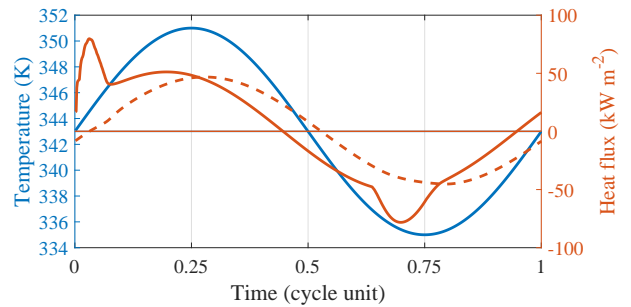}
\caption{\label{fig:fig_4}     Cyclic variations of the left thermal bath temperature (in blue), of the heat flux at left boundary of the PCM (continuous red line), and of the heat flux at right boundary (dashed red line), as obtained when the preliminary equilibrium temperature of 343~K is reached from below (case described in Fig. \ref{fig:fig_3}(a)). A thin horizontal red line depicts the net heat flux, here 0.124~kW$\,$m$^{-2}$, which yields a shuttling factor of 0.079\%.}
\end{figure}

Figure \ref{fig:fig_4} shows the cyclic evolution of the left bath temperature, and of the heat flux at both boundaries of the PCM for the same conditions as before while assuming the first prior scenario (the equilibrium temperature $T_{c}=343$~K is reached from below before starting the cycling of the left bath temperature). The results when assuming the second prior scenario (the equilibrium temperature $T_{c}=343$~K is reached from above), although significantly different, are indiscernible by eye, therefore they are not reproduced here. At the considered frequency, i.e. 0.3 Hz, the PCM layer is thermally "intermediate" (i.e. neither thin nor thick), this is the reason why the heat flux at the right boundary is neither equivalent to the one at the left boundary nor considerably lower. This also explains the phase shifts between the three curves. The heat flux at the right boundary is nearly sinusoidal, probably because the right part of the sample behaves linearly. In contrast, the heat flux at the left boundary shows a bump at the beginning of the first quarter and a depression at the end of the third quarter. Those are coincident with the two time intervals where the left part of the PCM undergoes the strongest changes in conductivity. 

The mean heat flux during one period (i.e. the net heat flux) is not zero, contrarily to what is observed with a linear material. It is however very small as compared to the magnitude of the extremal heat flux values on both left and right boundaries. We added in Fig. \ref{fig:fig_4} a horizontal line at the level of the net heat flux; it is hardly discernible with the baseline. Indeed, the net heat flux is only 0.124 kW$\,$m$^{-2}$ whereas the heat flux at the left boundary of the PCM oscillates between -78.2 and +79.8 kW$\,$m$^{-2}$. The flux modulation is not symmetric, therefore we propose the following definition for the heat-shuttling factor $S$:
\begin{equation}\label{eq:Shuttling_factor}
S=\frac{\overline{\varphi(t)}}{\varphi_{max}-\varphi_{min}},
\end{equation}
where $\varphi_{max}$ and $\varphi_{min}$ are the maximum, resp. minimum values of the conduction flux at the left boundary (i.e. where the temperature modulation is imposed) and $\overline{\varphi(t)}$ is the observed net heat flux (shuttling effect). The heat-shuttling factor is 0.079\% with the first prior scenario; it is just slightly higher, namely 0.081\%, with the second scenario, see Table \ref{tab:table_Shuttling}.
From the previous observations we can infer that the PCM can manifest its nonlinearity dynamically and hence give a chance to exhibit a heat-shuttling effect only if the left reservoir cools down sufficiently to reach the upper boundary of the cooling branch of the hysteresis loop, namely 337~K. The amplitude of the oscillations should thus be higher than 6~K. For lower amplitudes, the PCM behaves like a graded, yet linear, material (1\textsuperscript{st} prior scenario) or stays fully in the metallic phase (2\textsuperscript{nd} prior scenario).

A way to see a heat-shuttling effect for lower amplitude values of the left thermal bath is to move the mean temperature $T_{a}$ closer to the median temperature of the main hysteresis loop, i.e. 339~K. Actually, the previous statement can be generalized by saying that the PCM can exhibit a heat-shuttling effect only if $T_{L}(t)$ reaches the conductivity-variable-part \textit{of both} the cooling branch \textit{and} the heating branch, at least partially.

\begin{table}
\caption{\label{tab:table_Shuttling}
Characteristic data (minimum and maximum heat flux at the left boundary of the PCM, net heat flux and shuttling factor) obtained for two different values of the mean temperature $T_{c}$ and depending on whether the initial equilibrium temperature $T_{c}$ is reached from below of from above. Common parameters are: VO$_{2}$ sample thickness $l=1$~mm, frequency $f=0.3$~Hz, amplitude $T_{a}=8$~K. 
}
\begin{ruledtabular}
\begin{tabular}{cccccc}
\textrm{$T_{c}$}&
\textrm{Initial}&
\multicolumn{3}{c}{Flux (kW$\,$m$^{-2}$)}&
\textrm{Shuttling}\\
\textrm{(K)}&
\textrm{direction\footnote{Initial equilibrium temperature reached by heating ($\uparrow$) or cooling ($\downarrow$)}}&
\textrm{Min}&
\textrm{Max}&
\textrm{Net}&
\textrm{factor (\%)}\\
\colrule
343 &
$\uparrow$\footnote{see Fig. \ref{fig:fig_3}(a) and Fig. \ref{fig:fig_4}}  &
 -78.21 & +79.79 & +0.124 & 0.0787\\
343 & $\downarrow$\footnote{see Fig. \ref{fig:fig_3}(b)}  &
 -78.65 & +79.42 & +0.128 & 0.0812\\
339 &$\uparrow$\footnote{see Fig. \ref{fig:fig_5}(a) and Fig. \ref{fig:fig_6}(a)}&
 -125.6 & +141.2 & +1.00 & 0.377\\
339 &$\downarrow$ \footnote{see Fig. \ref{fig:fig_5}(b) and Fig. \ref{fig:fig_6}(b)}& 
-126.4 & +142.0 & +1.40 & 0.522\\
\end{tabular}
\end{ruledtabular}
\end{table}

\begin{figure}
\includegraphics[width=0.45\textwidth]{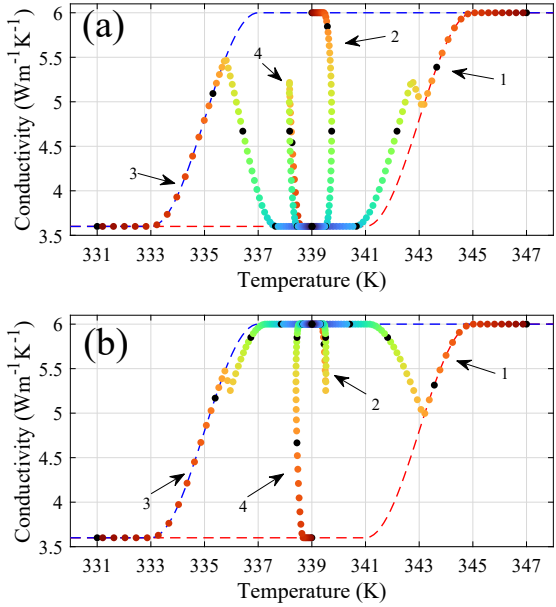}
\caption{\label{fig:fig_5}Same as in Fig. \ref{fig:fig_3} after changing $T_{c}$, the preliminary equilibrium temperature and the mean temperature of the baths to 339~K, which is the median temperature of the VO$_{2}$ hysteresis loop. See the videos of a whole cycle in the Supplementary Material in animation-fig-5a, resp. animation-fig-5b.}
\end{figure}

Let us now consider the situation where $T_{a}=339$~K, all things equal otherwise. The corresponding temperature distributions inside the PCM after 1, 2, 3 and 4 quarters of a period are reported in Fig.~\ref{fig:fig_5}(a) for the first prior scenario and in Fig.~\ref{fig:fig_5}(b) for the second prior scenario. The corresponding video animations are provided in the Supplementary Material. We can notice that 21 (resp. 18) cells on the left side of the PCM (about 40\%, resp. 34\%, of total thickness) experience conductivity changes, among them 6 (about 11\%) 
experience a full hysteresis loop. The remaining cells in the right part, which amounts to 60\%(resp. 66\%) of total thickness, see no conductivity change: 35\%( resp. 30\%) manifest flat loops anchored on the heating branch (resp. cooling branch) and 25\%(resp. 36\%) near the right reservoir stay at conductivity level of the insulator phase (resp. metallic phase) in the case of the 1\textsuperscript{st} prior scenario (resp. 2\textsuperscript{nd} prior scenario).

The temperature $T_{L}(t)$ and the fluxes $\varphi_{L}(t)$ and $\varphi_{R}(t)$ are reported in Fig. \ref{fig:fig_4} for both scenarios. Higher differences between the two scenarios are now observed, especially for the flux at the right boundary, $\varphi_{R}(t)$, mainly because in this area the material stays in two very different states, either in the insulator phase or in the metallic phase. The net heat flux was again added, which can now be distinguished from the baseline (altough with difficulty). As reported in Table \ref{tab:table_Shuttling} the heat-shuttling factor now reaches higher values, 0.38\% and 0.52\% with the 1\textsuperscript{st}, resp. 2\textsuperscript{nd} prior scenario. The observed increase, as compared to when $T_{a}$ was set to 339~K, is due to the fact that a larger portion of the layer is allowed to manifest the intrinsic nonlinearity in conductivity of the PCM material. 

\begin{figure}
\includegraphics[width=0.45\textwidth]{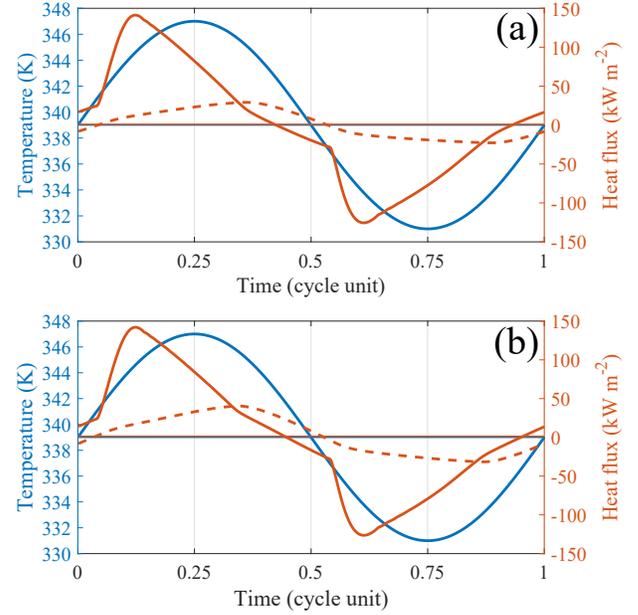}
\caption{\label{fig:fig_6}Same as in Fig. \ref{fig:fig_4} after changing $T_{c}$ to 339~K. In (a) the preliminary equilibrium temperature of 339~K is reached from below, in (b) it is reached from above. The net heat flux is 1.0 kW$\,$m$^{-2}$ in (a) (thin horizontal red line), and 1.4 kW$\,$m$^{-2}$ in (b), which yields a heat-shuttling factor of 0.38\%, resp. 0.52\%.}
\end{figure}

\subsection{\label{sec:Parametric}Parametric analysis}
A parametric analysis was performed in order to evaluate the sensitivity of the net heat flux to the PCM properties and operational conditions, and hence to find ways to increase the heat-shuttling factor.

\begin{figure} 
\includegraphics[width=0.45\textwidth]{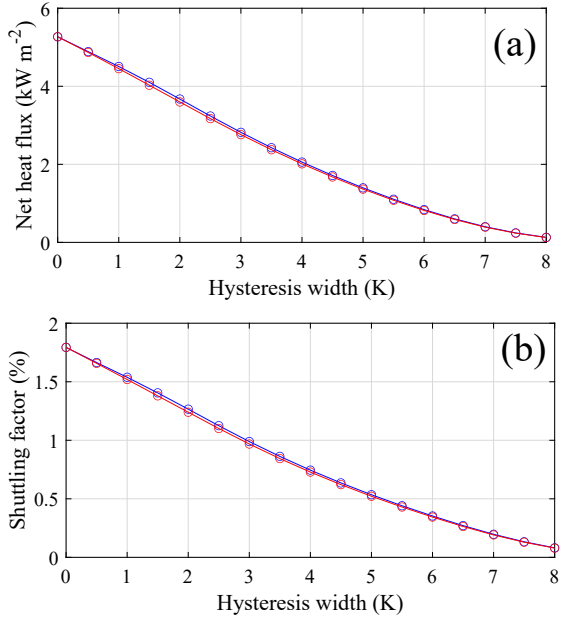}
\caption{\label{fig:fig_7}Net heat flux (a) and shuttling factor (b) as a function of the hysteresis width for the following conditions:  VO$_{2}$ sample thickness: 1 mm, frequency: 0.3 Hz, temperature amplitude: $T_{a}=8$~K, mean temperature: $T_{c}=343$~K. The preliminary equilibrium temperature $T_{c}$ is initially reached by heating (red curve) or by cooling (blue curve). }
\end{figure}

Figure \ref{fig:fig_7} 
describes the influence of the hysteresis width on the net heat flux and the heat-shuttling factor. The conditions are those of the first analysis in Fig. \ref{fig:fig_3}, \ref{fig:fig_4}, namely $T_{c}=343$~K, $T_{a}=8$~K, and $f=0.3$~Hz. The results show that an increase of the hysteresis width induces a rapid, almost linear, decrease of the heat-shuttling effect. The prior scenario has nearly no effect on the results. When comparing the real case of 8~K hysteresis width with the hypothetical case where there is no hysteresis, the net heat flux is lower by a factor of 0.024 in the real case and the heat-shuttling effect by a factor of 0.045.

\begin{figure}
\includegraphics[width=0.45\textwidth]{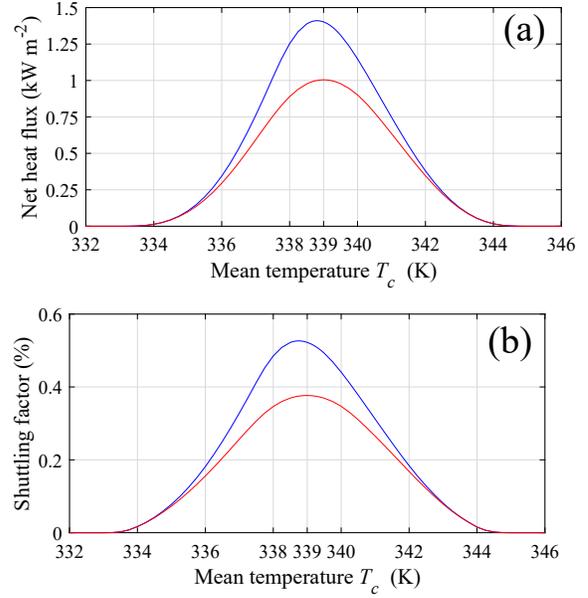}
\caption{\label{fig:fig_8}Net heat flux (a) and shuttling factor (b) as a function of the temperature mean value $T_{c}$ for the following conditions:  VO$_{2}$ sample thickness: 1 mm, frequency: 0.3 Hz, temperature amplitude: $T_{a}$=8~K. The preliminary equilibrium temperature $T_{c}$ is initially reached by heating (red curve) or by cooling (blue curve). }
\end{figure}

Figure \ref{fig:fig_8} 
describes the influence of the mean temperature of the PCM on, again, the net heat flux and the heat-shuttling factor. The other conditions are $T_{a}=8$~K, $f=0.3$~Hz, and a hysteresis width of 8~K. The net heat flux and the heat-shuttling factor present a maximum which corresponds, or is close to, the median temperature of the hysteresis main loop, i.e. 339~K. The results confirm what was already noticed before (see Table \ref{tab:table_Shuttling}), namely that the shuttling effect is stronger when applying the 2\textsuperscript{nd} prior scenario, i.e. by reaching the mean temperature from above before starting the cycling (there is approximately a 40\% increase in net heat flux and heat-shuttling factor when applying this scenario). One can also notice that with the 2\textsuperscript{nd} prior scenario, the optimal $T_{c}$ (338.8~K) is slightly lower than the median temperature of the hysteresis main loop (339~K).

\begin{figure}
\includegraphics[width=0.45\textwidth]{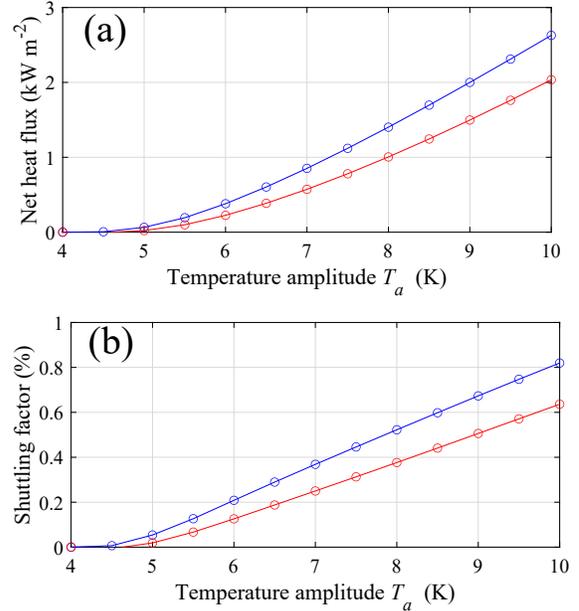}
\caption{\label{fig:fig_9}Net heat flux (a) and shuttling factor (b) as a function of the temperature modulation amplitude $T_{a}$ for the following conditions:  VO$_{2}$ sample thickness: 1 mm, frequency: 0.3 Hz, mean temperature: $T_{c}=339$~K. The preliminary equilibrium temperature $T_{c}$ is initially reached by heating (red curve) or by cooling (blue curve). }
\end{figure}

Figure \ref{fig:fig_9} 
describes the influence of the oscillation amplitude of the PCM. The other conditions are $T_{c}=339$~K, $f=0.3$~Hz, and a hysteresis width of 8~K. The net heat flux and the heat-shuttling factor monotonically increase with the amplitude of the temperature modulation of the left reservoir. They are higher and their rate of increase is higher as well when applying the 2\textsuperscript{nd} prior scenario.

\begin{figure}
\includegraphics[width=0.45\textwidth]{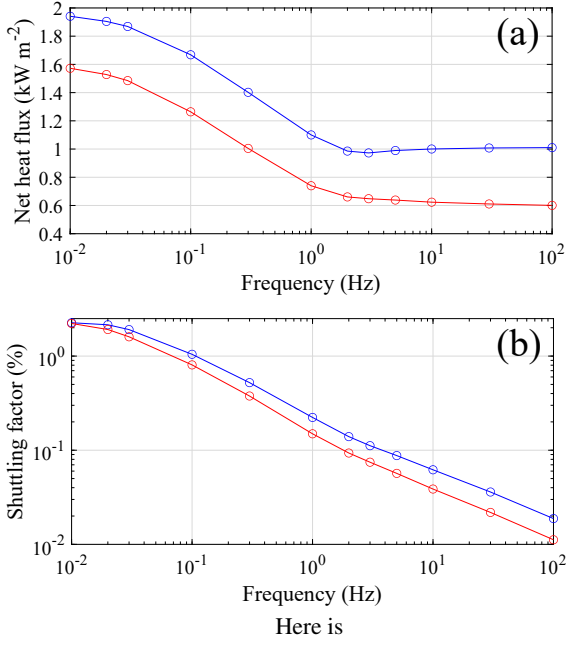} Here is 
\caption{\label{fig:fig_10}Net heat flux (a) and shuttling factor (b) as a function of the frequency of the temperature modulation of the left bath for the following conditions:  VO$_{2}$ sample thickness: 1 mm, temperature amplitude: $T_{a}=8$~K, mean temperature: $T_{c}=339$~K. The preliminary equilibrium temperature $T_{c}$ is initially reached by heating (red curve) or by cooling (blue curve). }
\end{figure}

Figure \ref{fig:fig_10} 
describes the influence of the modulation frequency. The other conditions are $T_{c}=339$~K, $T_{a}=8$~K, a hysteresis width of 8~K and a VO$_{2}$ layer 1 mm thick. Whatever the value of the frequency, the net heat flux is always obtained higher (from 20\% to 70\%) with the 2\textsuperscript{nd} prior scenario. The net heat flux is highest at vanishing frequency (thermally thin PCM case). After a slow decrease in the thermally thin range ($f \in$[0, 0.03]~Hz), the net heat flux decreases more rapidly in the intermediate range ($f \in$[0.03, 1]~Hz), and then, depending on the scenario, either it shows a minimum at  $\sim$0.97 kW$\,$m$^{-2}$ for a frequency of about 3~Hz, and finally reaches a plateau at about 1 kW$\,$m$^{-2}$ (2\textsuperscript{nd} prior scenario), or it continues to decrease but at a slower rate (1\textsuperscript{st} prior scenario). Because the amplitude of the heat-flux excursions on the PCM left boundary continuously increases with the frequency, the heat-shuttling factor monotonically decreases with the frequency, whatever the scenario. Starting from values close to 2.2\% at 0.01~Hz (nearly static regime) the heat-shuttling factor first slowly decreases to 0.8-1\% at 0.1~Hz, then decreases faster as $f^{\alpha}$ with $\alpha$ between $-2/3$ and $-1/2$.

\subsection{\label{sec:two-bath-modulation}Heat-shuttling with two-bath modulation}
\begin{figure}
\includegraphics[width=0.45\textwidth]{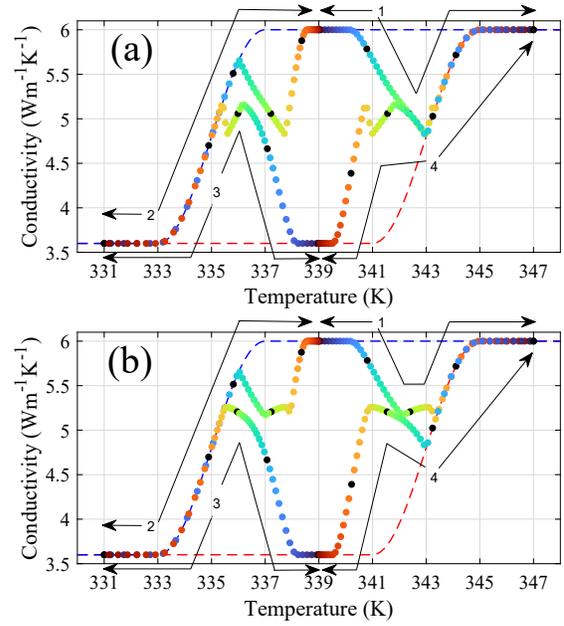}
\caption{\label{fig:fig_11}Same as in Fig. \ref{fig:fig_3} when both reservoirs are modulated with the same mean temperature $T_{c}=339$~K, same amplitude $T_{a}=8$~K, same frequency $f=0.3$~Hz, and a phase lag $\delta=3\pi/2$ (the right bath is in quadrature ahead). In (a) the first prior scenario was applied ($T_{c}$ reached from below) whereas in (b) the second prior scenario was applied ($T_{c}$ reached from above). See the videos of a whole cycle in the Supplementary Material in animation-fig-11a, resp. animation-fig-11b.}
\end{figure}

\begin{figure}
\includegraphics[width=0.45\textwidth]{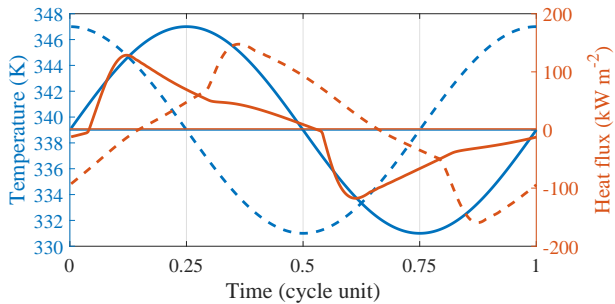}
\caption{\label{fig:fig_12}Cyclic evolution of temperature (in blue) and heat flux (in red) at the left reservoir (continuous lines), and right reservoir (dashed lines) in the conditions of Fig. \ref{fig:fig_11} when applying the second prior scenario (the curves for the first scenario are visually the same). A thin red horizontal line indicates the neat heat flux level.}
\end{figure}

\begin{figure}
\includegraphics[width=0.45\textwidth]{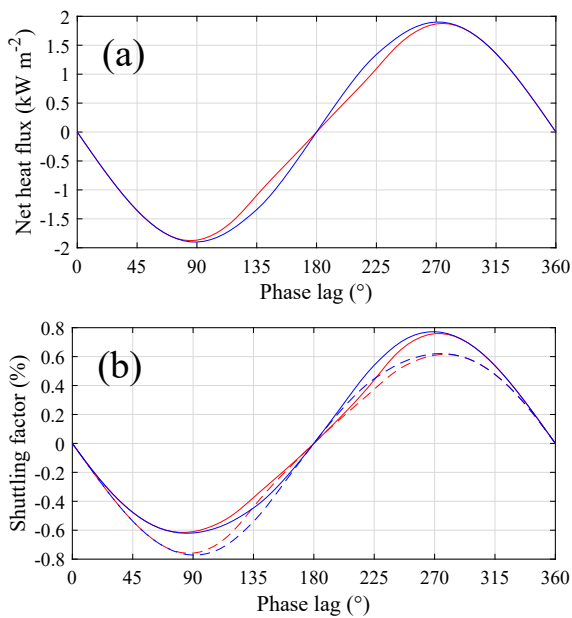}
\caption{\label{fig:fig_13}Net heat flux (a) and shuttling factor (b) as a function of the phase lag $\delta$ (in degrees) between the two reservoirs for the following conditions:  VO$_{2}$ sample thickness: 1 mm, temperature amplitude: $T_{a}=8$~K, mean temperature: $T_{c}=339$~K, frequency: $f=0.3$~Hz. The preliminary equilibrium temperature $T_{c}$ is initially reached by heating (red curve) or by cooling (blue curve). The shuttling factor is with respect to the flux range on the left side (continuous lines) or the right side (dashed lines).}
\end{figure}

Until now, the heat shuttling was produced by temperature oscillations of one reservoir, the other being maintained at the average temperature of the first. We will now explore the case where both reservoirs experience temperature modulations while maintaining a zero mean thermal bias. The analysis will be focused on the process where the thermal baths share the same temperature modulation (same mean $T_{c}$, same amplitude $T_{a}$ and same frequency $f$) but with a phase lag $\delta$:
 \begin{equation}\label{eq:BC2_l}
T_{L}(t)=T_{c}+T_{a} \sin\left( 2\pi f t \right) ,
\end{equation}
 \begin{equation}\label{eq:BC2_r}
T_{R}(t)=T_{c}+T_{a} \sin\left( \max\left(0,2\pi f t -\delta \right)\right),
\end{equation}
with $\delta$ in $[0,2\pi]$. According to Eq.~(\ref{eq:BC2_l}),(\ref{eq:BC2_r}), both thermal baths are at temperature $T_{c}$ at $t=0$ which is assumed to be the initial temperature of the PCM as well. As before, this initial state is considered to have been reached either by heating or by cooling. Subsequently, the temperatures are continuously modulated with the right bath behind the left one when $\delta\in[0,\pi]$, or ahead when $\delta\in[\pi,2\pi]$.
Figure~\ref{fig:fig_11} describes the temperature/conductivity distribution inside the PCM at four different times of a cycle when the temperatures of the reservoirs are in quadrature. Some differences can be noticed depending on whether one applies the first or second prior scenario, but they essentially concern the core of the material. Temperature and heat flux on the boundaries of the PCM are reported as a function of time in Fig.~\ref{fig:fig_12} for the prior scenario where the mean temperature $T_{c}=339$~K is previously reached from above. The other case leads to substantially the same curves; in particular, the net heat flux reaches very close values, 1.90 and 1.87 kW$\,$m$^{-2}$ for the 1\textsuperscript{st}, resp. 2\textsuperscript{nd} scenario. The results for different values of the phase lag are summarized in Fig.~\ref{fig:fig_13}
regarding the net heat flux and the shuttling factor. This factor can be defined in two ways depending on whether the heat flux range at the denominator of Eq.~(\ref{eq:Shuttling_factor}) is taken on the left or right boundary. Interestingly, the direction of the net heat flux depends on the phase lag. It is always towards the reservoir whose modulation is ahead. In addition it vanishes when the modulations are in phase or out of phase. A maximum for the net heat flux is observed for a phase lag $\delta$ close to $3\pi/2$ (a maximum in the opposite direction is observed for a phase lag close to $\pi/2$). When compared to the case where only the left reservoir is modulated, the net heat flux has increased by 90\% (from 1.00 to 1.87 kW$\,$m$^{-2}$) with the 1\textsuperscript{st} prior scenario, and by 34\% (from 1.40 to 1.90 kW$\,$m$^{-2}$) with the 2\textsuperscript{nd} prior scenario. The shuttling factor (when considering the flux range at the left edge) respectively increases from 0.377\% to 0.759\% and from 0.522\% to 0.772\%. Despite this relative improvement, the shuttling factor remains lower than 1\%.

\section{\label{sec:Discussion}Discussion and Conclusion}
It was demonstrated in \cite{ordonez2022net} that in a nonlinear material submitted to a quasi-static temperature modulation on one side, a significant heat-shuttling effect can be observed. A theoretical illustration was provided with VO$_{2}$ which, because of a MIT, shows a strong variation of conductivity (+67\%) when temperature increases from values lower than $\sim333$~K (insulating phase) to values higher than $\sim345$~K. The thermal hysteresis of VO$_{2}$ was ignored in \cite{ordonez2022net} and vanishingly slow temperature modulations (quasi-static regime) were assumed. Anywhere inside the material, the oscillations of the representative point $\left\lbrace T,\,k\right\rbrace $ were restrained to the heating branch of the main hysteresis loop of VO$_{2}$. As an example, for a temperature modulation of 5~K, a net heat flux of 2.9 kW$\,$m$^{-2}$ and a heat-shuttling factor of about 5.8\% were obtained by setting the mean temperature at the VO$_{2}$ transition temperature along the heating branch, a condition that was found to yield the strongest heat-shuttling effect.

The heat-shuttling effect in PCM materials was also studied in  \cite{liu2022energy}, but introducing rough approximations regarding the thermal hysteresis, both for conductivity and apparent specific heat : when reversing the direction of the temperature evolution, an instantaneous switch between the heating and cooling branches was applied, but no indication was provided regarding the shape of the partial hysteresis loops. It was then shown that increasing the hysteresis width (by keeping it lower than the temperature modulation amplitude) induced a reduction of the net heat flux by a few tens of percent. The considered values for frequency and PCM thickness corresponded to the thermally thick regime.

In this paper we explored the influence of the PCM hysteresis and modulation frequency on the heat-shuttling effect by modeling carefully the coupled variations of temperature, conductivity and enthalpy, especially for alternating episodes of cooling or heating of low amplitude. Thereby we transposed the concept of non hysteretic branches (NHBs) observed for electric resistance and optical reflectance of VO$_{2}$ \cite{ramirez2009first,gurvitch2009nonhysteretic}
to thermal conductivity and enthalpy, while extrapolating it to wider branches, up to the width of the whole hysteresis loop, which, regarding enthalpy, corresponds to the model described in \cite{bony2007numerical} and which could be called the \emph{gradual curve switching method}.

In summary, we confirm the existence of a shuttling of heat by conduction in a nonlinear material under nonequilibrium conditions. However, we found that hysteresis, when present, contributes to reduce to a large extent the net heat flux, moreover it makes that the heat shuttling effect is dependent on the past thermal history, i.e. before starting the temperature modulation of the reservoir. The reduction of the net heat flux is due to the appearance of NHBs inside the material. In fact, since the material undergoes a temperature modulation of decreasing amplitude as one approaches the static thermal bath, the hysteresis loops gradually thin, then flatten and become NHBs. In the vicinity of the static bath where the conductivity hysteresis loops have flattened and transformed to NHBs, the conductivity remains \emph{static} (although possibly variable in space), hence the potential nonlinearity $k(T)$ does not show up locally. As a consequence, the part of the PCM layer close to the static reservoir does not participate to the heat-shuttling effect (this is valid in the thermally thin regime - low to vanishing frequency -  and the thickness of the non-participating part increases with frequency). In addition, the temperature amplitude of the modulated bath must exceed a value depending on the global mean temperature and on the hysteresis width for the heat-shuttling effect to appear. As an example, when setting the global mean temperature at the transition temperature along the heating branch (i.e. 343~K), the modulation amplitude must exceed 6~K, which is higher than the values considered in \cite{ordonez2022net}. Even with an amplitude of 8~K, the heat-shuttling factor remains at low values, whatever the prior scenario, namely less than 0.1\% in the case of a modulation frequency of 0.3~Hz and a 1 mm thick VO$_{2}$ layer. The net heat flux and the shuttling factor can be maximized by setting the global mean temperature close to the median temperature of the VO$_{2}$ main hysteresis loop (the exact value depends on the prior scenario). Increasing the modulation amplitude beyond a minimum threshold helps to amplify the heat-shuttling effect. Another mean is to reduce the modulation frequency, down to the thermally thin regime. Among the two prior scenarios considered, i.e. with a preliminary equilibrium temperature reached by heating or cooling, we have shown that the second scenario systematically produces a stronger heat-shuttling effect.

Nevertheless it should be noted that the phenomenon remains low in intensity when considering VO$_{2}$. For example, with a temperature amplitude of 8~K, the shuttling factor is no more than about 2\% in the quasi-static regime (thermally thin VO$_{2}$ layer) and it drops rapidly with frequency (e.g.  $\sim$0.2\% at 1~Hz for a 1 mm thick layer). The net heat flux therefore reaches only a very small fraction of the flux amplitude observed at the modulated reservoir. However, there remains  the possibility of leveraging the sensitivity of the reported shuttling effect to the amplitude of the periodic thermal excitation in order to increase the shuttling factor somewhat.

Notwithstanding, our work offers a useful tool for predicting and interpreting the response of a dynamical thermal system encompassing PCMs with hysteresis. A potential extension is the inclusion of radiative effects whereby emissivity is nonlinear and presents hysteresis as well, with the objective of an analysis of the interplay between nonlinear conduction, heat storage, thermal emission, all presenting hysteresis, for an improved modeling of thermal diodes, thermal transistors, thermal logic gates, thermal memories, and thermal memristors.

\section{\label{sec:SupplMat}Supplementary material}
See the supplementary material for the details of the numerical model, a verification performed against an analytical model for the limiting case of a linear material, and a series of six animations provided in mp4 format showing the dynamic evolution of joint temperature and conductivity distributions corresponding to Fig.~\ref{fig:fig_3}, \ref{fig:fig_5}, and \ref{fig:fig_11} .

\bibliography{Shuttling_Biblio_230116}

\end{document}